\documentclass[aip,jcp,reprint,onecolumn,showpacs,showkeys]{revtex4-1}
\usepackage{graphics}
\usepackage{graphicx}
\usepackage{epsfig}
\usepackage{amssymb}
\usepackage{amsthm}
\usepackage{amsmath}
\usepackage{amstext}
\usepackage{natbib}

\begin{document}
\title{The Hydrophobic Aggregation of Two Colloids: A Thermodynamic Model}
\author{Pierre \surname{de Thier}}
\email{pierre.dethier@uclouvain.be}
\affiliation{Universit\'{e} catholique de Louvain, Louvain-la-Neuve (Belgium)}
\date{\today}

\begin{abstract}
Colloidal aggregation could be implemented in various fields ranging from purely colloidal thermodynamics to protein interactions, their stability, and maybe folding. Indeed, colloidal aggregation is closely linked to the so-called hydrophobic effect for which a thermodynamic explanation is proposed. This explanation is performed using Prigogine's out-of-equilibrium thermodynamics which is based on entropy production. It is shown that a likely destabilizing event could induce a spontaneous and irreversible aggregation of two identical or different colloids as it causes the desorption of solvent molecules from colloidal surfaces. This desorption is an entropy production factor through the chemical potentials minimization of, among others, initially adsorbed molecules. This may be viewed as an increase of the so-called ``solvent entropy'' or ``translational entropy''.
\end{abstract}

\maketitle

\section*{Introduction}
Since biomolecules, like proteins and DNA are colloids, the process of colloidal assembly seems to be a major field of interest. Indeed, understanding how colloids aggregate may provide an interesting conceptual framework for biomolecular self-assemblies and interactions between these biomolecules. Colloidal assemblies are only one aspect of the very multifaceted hydrophobic effect \citep{chandler2005}. In an important paper on the hydrophobic effect and the organization of living matter, \citet{tanford1978} wrote that the biological organization may be viewed as consisting of two stages: the biosynthesis of appropriate molecules and the assembly of these molecules into organized structures. As \citet{tanford1978} explainds, life needs water for two reasons; first, water is an effective solvent to synthesize biomolecules and second, it induces these molecules to assemble into larger structures. Thus, suspending environment (water) appears to be more than simply a chemical medium but also a real matrix causing self-assembly, aggregation or folding \citep{levy2006}.

This type of assembly process could be applied to protein tertiary structures aggregation (\textit{e.g.} binding of ribosomal subunits with mRNA during the initiation stage of protein biosynthesis), ligand-receptor docking (\textit{e.g.} antigen-antibody interaction), and maybe also to protein folding \citep{lehninger,levy2006}. These phenomena must involve forces able to bind and specifically stick together the different subunits or parts of the same macromolecule in the case of protein folding. Today, it is well accepted that instead of one force, it seems that there rather is a sum of forces emanating from various origins and ensuing on the so-called hydrophobic effect \citep{dill2012}.

From general knowledge, hydrophobic effect is associated with uncharged organic molecules, van der Waals forces between them, and entropy \citep{brady1997}. It is not exactly known how they bring together in a hypothetical hydrophobic ``force'' but there are enough good reasons to believe in their implications \citep{lum1999,chandler2005}. Electrostatic charges and van der Waals forces are classified into kinetic considerations because they can form potential energy barriers steering the stability of colloids as shown by the DLVO model \citep{verwey1948}. In an other way, entropy arises from thermodynamic considerations and specifically the second law. In fact, in his famous book ``What is Life'', \citet{schrodinger1967} asserts that an organism feeds upon negative entropy. In other words, organisms live because they produce and then export their entropy (negentropy). In addition, considering entropy allow us to follow the evolution of a system during an out-of-equilibrium process. The second law gives the mandatory direction of this evolution: the evolution followed by the system toward equilibrium. Entropy production has been showed by Planck as the natural evolution criterion for a given system \citep{planck1913,guggenheim1949}.

Initially, let us consider colloids in a stable state. Then the system transforms because stability conditions have been broken. In this subsequently non-equilibrium and unstable situation, the system evolves following thermodynamics which has been extensively developed by Prigogine \textit{et al}. \citep{defay1977,prigogine1968}. It will evolve and produce entropy until it reaches the true thermodynamic equilibrium where chemical potentials of involved chemical species are equivalent (Gibbs' definition of equilibrium \citep{reiss1965}).

\begin{figure}
\begin{center}
\includegraphics[angle=0,width=8cm,height=2.683724236cm]{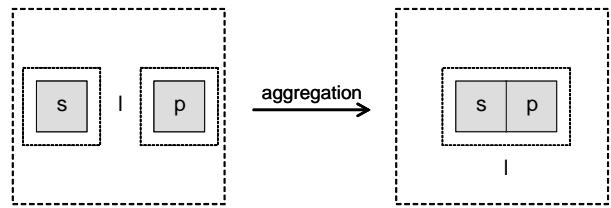}
\end{center}
\caption{Initial and final states of an aggregating colloidal system. Two colloids ``s'' and ``p'' are aggregating into a liquid phase ``l''. Those two colloids are surrounded by superficial phases ``sl'' and ``pl'' coarsely represented by dotted belts around them. Solid and dashed lines represent boundaries of various phases. They are dashed for open phases (energy and matter exchanges) and solid for closed ones (no matter exchange). System is delimited by the external dashed line because it is open to its environment.}\label{fig}
\end{figure}

In this paper, I would like to propose some equations describing the aggregation process of two colloids through hydrophobic effect. These equations will be inferred from the irreversible thermodynamic framework of Prigogine \textit{et al}. \citep{defay1977,prigogine1968}. As illustrated on figure \ref{fig}, this will be done according to a very simple scheme of aggregation of two colloids ``s'' and ``p'' suspended in a liquid phase ``l''. In the first section, well known equations for energy conservation based on the first law of thermodynamics will be developed. Then, in the second section, the second law will be introduced and the entropy production will allow us to obtain equations  remaining valid for out-of-equilibrium situations. The equations which will be obtained will allow us to discuss about equilibrium conditions and stability conditions which are not necessarily the same. Since the equilibrium conditions are based on chemical potentials, stability conditions involve other forces such as electrostatic and van der Waals.

It is worth noting that the following reasoning considers that colloids are real phases (or mesophases \citep{chandler2005}) such that the system may be divided into several equations for each of these phases \citep{guggenheim1949}. By contrasting with other studies \citep{nagarajan1991,tanford1978}, considering colloids as real phases also implies that we can not define a chemical potential for them at all whatever their nature (proteins, polymers, \textit{etc}.).

\section{Energy Conservation Along the Aggregation Process: Application of the First Law}\label{FirstSection}

The first law of thermodynamics describes the conservation of energy through the concept of internal energy $E$. It may be enunciated as follows \citep{guggenheim1949,prigogine1968,prigogine1950}:
\begin{equation}\label{FirstLaw}
{\rm{d}}E={\rm{d}}Q+{\rm{d}}W.
\end{equation}
Equation (\ref{FirstLaw}) shows that the variation of the internal energy ${\rm{d}}E$ of the whole system is equal to ${\rm{d}}Q$ the heat which has been given by the environment to the system during the transformation and ${\rm{d}}W$ the work done on it during the same period of time ${\rm{d}}t$. In our case, this work may be of two kinds: expansionary and chemical. The expansion work depends on a change in volume ${\rm{d}}v$ of the whole system and expands
\begin{equation}
{\rm{d}}W=-p^{\rm{l}}{\rm{d}}v
\end{equation}
where $p^{\rm{l}}$ is the pressure in the liquid phase. This last expression must be substituted into equation (\ref{FirstLaw}) to give internal energy variation for a closed system.

\begin{widetext}
In a capillary system containing various bulk and superficial phases, the mechanical work can be transformed to give some useful expressions. First, we can use the fact that $v=v^{\rm{l}}+v^{\rm{p}}+v^{\rm{s}}$ to obtain equation (\ref{Work1}) \citep{defay1951}.
\begin{equation}\label{Work1}
{\rm{d}}W=-p^{\rm{l}}{\rm{d}}v^{\rm{l}}-p^{\rm{l}}{\rm{d}}v^{\rm{p}}-p^{\rm{l}}{\rm{d}}v^{\rm{s}}
\end{equation}
Let us now transform equation (\ref{Work1}) into
\begin{equation}\label{Work2}
{\rm{d}}W=
-p^{\rm{l}}{\rm{d}}v^{\rm{l}}
-p^{\rm{l}}{\rm{d}}v^{\rm{p}}
-p^{\rm{l}}{\rm{d}}v^{\rm{s}}
-p^{\rm{p}}{\rm{d}}v^{\rm{p}}
+p^{\rm{p}}{\rm{d}}v^{\rm{p}}
-p^{\rm{s}}{\rm{d}}v^{\rm{s}}
+p^{\rm{s}}{\rm{d}}v^{\rm{s}}
\end{equation}
which is rearranged to give equation (\ref{Work3}).
\begin{equation}\label{Work3}
{\rm{d}}W=
-p^{\rm{l}}{\rm{d}}v^{\rm{l}}
-p^{\rm{p}}{\rm{d}}v^{\rm{p}}
-p^{\rm{s}}{\rm{d}}v^{\rm{s}}
+(p^{\rm{p}}-p^{\rm{l}}){\rm{d}}v^{\rm{p}}
+(p^{\rm{s}}-p^{\rm{l}}){\rm{d}}v^{\rm{s}}
\end{equation}
\end{widetext}

For spherical phases ``s'' and ``p'', the Laplace equation \citep{defay1951,hunter1987} gives
\begin{equation}
p^{\rm{p}}-p^{\rm{l}}=\gamma^{\rm{pl}}\frac{2}{R^{\rm{p}}}\quad\text{and}\quad
p^{\rm{s}}-p^{\rm{l}}=\gamma^{\rm{sl}}\frac{2}{R^{\rm{s}}}
\end{equation}
where $\gamma^{\rm{pl}}$ and $\gamma^{\rm{sl}}$ are respectively superficial tensions for interfaces between phases ``p'' and ``l'', and between phases ``s'' and ``l''. $R^{\rm{p}}$ and $R^{\rm{s}}$ are radius of the two colloidal phases ``p'' and ``s''. After injection of these results and knowing that
\begin{equation}
{\rm{d}}\sigma=\frac{2}{R}{\rm{d}}v
\end{equation}
for spheres ($\sigma$ is the surface of the sphere), we obtain equation (\ref{Work4})
\begin{equation}\label{Work4}
{\rm{d}}W=
-p^{\rm{l}}{\rm{d}}v^{\rm{l}}
-p^{\rm{p}}{\rm{d}}v^{\rm{p}}
-p^{\rm{s}}{\rm{d}}v^{\rm{s}}
+\gamma^{\rm{pl}}{\rm{d}}\sigma^{\rm{pl}}
+\gamma^{\rm{sl}}{\rm{d}}\sigma^{\rm{sl}},
\end{equation}
a new expression for expansion work. Equation (\ref{Work4}) is more general as there is no need of geometrical considerations and thus also applies to planar surfaces \citep{defay1951}. Now, our interest is not really expansion but a specific change: aggregation between colloidal phases ``s'' and ``p''. If supposing that volumes are exactly counterbalanced such that
\begin{equation}\label{VolumesNuls}
-p^{\rm{l}}{\rm{d}}v^{\rm{l}}
-p^{\rm{p}}{\rm{d}}v^{\rm{p}}
-p^{\rm{s}}{\rm{d}}v^{\rm{s}}=0
\end{equation}
and that the aggregation phenomena results in the creation of a new interface between phases ``s'' and ``p'' associated with a surface tension $\gamma^{\rm{sp}}$ and an area $\sigma^{\rm{sp}}$, it follows from (\ref{Work4}) and (\ref{VolumesNuls}):
\begin{equation}\label{Work5}
{\rm{d}}W=
\gamma^{\rm{pl}}{\rm{d}}\sigma^{\rm{pl}}
+\gamma^{\rm{sl}}{\rm{d}}\sigma^{\rm{sl}}
+\gamma^{\rm{sp}}{\rm{d}}\sigma^{\rm{sp}}.
\end{equation}
If aggregation destroys as much ``pl'' and ``sl'' interfaces separately as it creates ``sp'' interface \citep{hunter1987}, then it comes
\begin{equation}\label{EquivSurfaces}
{\rm{d}}\sigma^{\rm{pl}}={\rm{d}}\sigma^{\rm{sl}}=-{\rm{d}}\sigma^{\rm{sp}}
\end{equation}
which after substitution into (\ref{Work5}) and factorization, provides the adhesion or aggregation work also known as Dupr\'{e} equation \citep{hunter1987}:
\begin{equation}\label{Work6}
{\rm{d}}W=(\gamma^{\rm{sp}}-\gamma^{\rm{sl}}-\gamma^{\rm{pl}}){\rm{d}}\sigma^{\rm{sp}}.
\end{equation}
Injection of Dupr\'{e} equation (\ref{Work6}) into the first law (\ref{FirstLaw}) gives the variation of internal energy during an aggregation process inside a closed system for which expansion works of each phase counterbalance.

When a system can exchange matter with its environment, it is an open system. Therefore, it is necessary to add a term to the mechanical work to give equation (\ref{FirstLaw3}) \citep{prigogine1968}.
\begin{equation}\label{FirstLaw3}
{\rm{d}}E={\rm{d}}Q+(\gamma^{\rm{sp}}-\gamma^{\rm{sl}}-\gamma^{\rm{pl}}){\rm{d}}\sigma^{\rm{sp}}+\sum_j\tilde{\mu}_j^{\rm{l}}{\rm{d}}_en_j^{\rm{l}}
\end{equation}
The amount of matter exchanged during the transformation of the system is included in equation (\ref{FirstLaw3}) through ${\rm{d}}_en_j^{\rm{l}}$ the variation of the number of mole of chemical species $j$ in the liquid phase ``l''. The notation ``${\rm{d}}_e$'' stands for a variation due to an exchange with the environment. Each amount of exchanged matter carries some energy which takes the form of a chemical potential $\mu_j^{\rm{l}}$, the variation of internal energy in function of the quantity $n_j$ when heat (or entropy) and volume remain constant \footnote{It is customary to define chemical potential as the partial derivative of free enthalpy at $p$ and $T$ constant. Here, for the sake of simplicity, it is better to use the derivative of $E$ keeping $S$, $v$, and $\sigma$ constant.}. Here, the notation $\tilde{\mu}_j^{\rm{l}}$ allows to generalize the chemical potential $\mu_j^{\rm{l}}$ as an electrochemical potential written \citep{lyklema1}
\begin{equation}\label{ChemicalPotential}
\tilde{\mu}_j^{\rm{l}}=\big[\mu_j^\circ+RT\ln a_j^{\rm{l}}\big]_\mathrm{chem}+\big[z_j\mathcal{F}\psi^{\rm{l}}\big]_\mathrm{elec},
\end{equation} 
an equation in which contributions of purely chemical and electrostatic considerations have been distinguished. In equation (\ref{ChemicalPotential}), $\mu_j^\circ$ is the chemical potential when standard conditions of temperature and pressure are applied, $R$ the gas constant, $T$ the temperature, $a_j^{\rm{l}}$ the activity of chemical species $j$ in the liquid phase, $z_j$ the electric charge, $\mathcal{F}$ the Faraday, and $\psi^{\rm{l}}$ the electric potential in the liquid phase.

Electrochemical potential is the sum of $\mu_j^{\rm{l}}$, the true chemical potential, and of $z_j\mathcal{F}\psi^{\rm{l}}$ the electrostatic effect. When dealing with colloidal surfaces, it is important to consider $z_j\mathcal{F}\psi^{\rm{l}}$ because these surfaces may be more or less charged, thereby affecting the quantity of ions situated near the surface. Indeed, to fulfil Gibbs' equilibrium conditions, electrochemical potentials of the species $j$ must be equivalent everywhere in the system, so we have the equality between species located in the bulk and superficial phases \citep{reiss1965}:
\begin{equation}\label{GibbsEquili}
\tilde{\mu}_j^{\rm{l}}=\dots =\tilde{\mu}_j^{\rm{sl}}=\dots =\tilde{\mu}_j^{\rm{pl}}.
\end{equation}
The consequence of this equality is that the quantities of chemical species will be discontinuous near the interfaces to satisfy equilibrium conditions. These spacial heterogeneities near the surfaces lead to the concept of adsorption introduced by Gibbs and for which valuable discussion is given in the literature \citep{defay1951,hunter1987,reiss1965}. Adsorptions of $j$ at ``pl'' and ``sl'' interfaces are respectively defined so that
\begin{equation}\label{DefAdso}
\Gamma_j^{\rm{pl}}=\frac{n_j^{\rm{pl}}}{\sigma^{\rm{pl}}}
\quad\text{and}\quad
\Gamma_j^{\rm{sl}}=\frac{n_j^{\rm{sl}}}{\sigma^{\rm{sl}}}.
\end{equation}
Adsorptions are thermodynamic quantities and must not be confused with the molecular content of the Stern layer (or inner Helmholtz plane). Indeed, this description from the double-layer model is not thermodynamic.

\section{Aggregation Produces Entropy: Application of the Second Law}\label{SecondSection}

The second law asserts that every transformation in which the sole result is the absorption of heat from a reservoir and its complete conversion in a work are totally impossible \citep{atkins}. This means there is necessarily a loss of heat during the transformation, taking the shape of a heat dissipation: ${\rm{d}}Q^\prime$. Clausius' (in)equality \citep{prigogine1950} can be expressed as
\begin{equation}
{\rm{d}}S\geqslant\frac{{\rm{d}}Q}{T}\quad\text{and}\quad{\rm{d}}S=\frac{{\rm{d}}Q}{T}+\frac{{\rm{d}}Q^\prime}{T}
\end{equation}
when the uncompensated heat ${\rm{d}}Q^\prime$ is taken into account. ${\rm{d}}Q$ is an exchange of heat (\textit{cf}. equation (\ref{FirstLaw})) between the system and its environment, ${\rm{d}}Q^\prime$ a dissipated one, and ${\rm{d}}S$ is the total variation of entropy. Assuming that ${\rm{d}}_eS={\rm{d}}Q/T$ the flow of entropy and ${\rm{d}}_iS={\rm{d}}Q^\prime/T$ the dissipated one or entropy production, the last equation is rewritten as \citep{prigogine1950,lebon2008a}
\begin{equation}
{\rm{d}}S={\rm{d}}_eS+{\rm{d}}_iS.
\end{equation}
As the second law requests that ${\rm{d}}Q^\prime\geqslant0$ for a natural (irreversible and spontaneous) transformation, ${\rm{d}}_iS$ must also be positive (${\rm{d}}_iS\geqslant0$). In the limit case, it is zero when transformation is reversible (${\rm{d}}_iS=0$). Thus, the entropy production gives a direction to natural processes and find an expression for ${\rm{d}}_iS$ should give an evolution criterion for the system under study.

\begin{widetext}
Injecting of ${\rm{d}}Q=T{\rm{d}}_eS$ and substituting ${\rm{d}}_eS={\rm{d}}S-{\rm{d}}_iS$ into equation (\ref{FirstLaw3}) results after rearrangement in
\begin{equation}\label{Entropy2}
{\rm{d}}S=\frac{1}{T}{\rm{d}}E-\frac{1}{T}(\gamma^{\rm{sp}}-\gamma^{\rm{sl}}-\gamma^{\rm{pl}}){\rm{d}}\sigma^{\rm{sp}}
-\frac{1}{T}\sum_j\tilde{\mu}_j^{\rm{l}}{\rm{d}}_en_j^{\rm{l}}+{\rm{d}}_iS.
\end{equation}
Equation (\ref{Entropy2}) is a first relation for the total variation of entropy during aggregation \citep{prigogine1968}, giving an explicit development for ${\rm{d}}_eS$ (the three first terms of the right member) but not for ${\rm{d}}_iS$ the evolution criterion in the sense of the second law. To find an equation for ${\rm{d}}_iS$, let us consider the system composed of different parts: three $\alpha$ bulk phases ``p'', ``s'', and ``l'' and three $\beta$ superficial phases ``sl'', ``pl'', and ``ps''. If for each separate phase the transformation remains reversible, that is ${\rm{d}}S={\rm{d}}_eS$ (${\rm{d}}_iS=0$), it is possible to obtain the entropy variation of the whole system ${\rm{d}}S$ by adding all ${\rm{d}}S$ obtained for each part (each $\alpha$ and $\beta$ phases composing the system) \citep{prigogine1968,guggenheim1949}. Therefore, equation (\ref{Entropy3}) comes as follow
\begin{equation}\label{Entropy3}
{\rm{d}}S=\frac{1}{T}\sum_{\alpha,\beta}{\rm{d}}E^{\alpha,\beta}+\frac{1}{T}\sum_{\alpha} p^{\alpha} {\rm{d}}v^{\alpha}-\frac{1}{T}\sum_{\beta}\gamma^{\beta}{\rm{d}}\sigma^{\beta}-\frac{1}{T}\sum_{\alpha,\beta}\sum_j\tilde{\mu}_j^{\alpha,\beta}{\rm{d}}n_j^{\alpha,\beta},
\end{equation}
an equation in which summations can be made on all $\alpha$ and $\beta$ phases. The first term in the right part of equation (\ref{Entropy3}) simplifies into ${\rm{d}}E$, the variation of internal energy of the whole system. The second term is zero as a consequence of equation (\ref{VolumesNuls}) and the third is the adhesion or aggregation work from equation (\ref{Work6}). So, equation (\ref{Entropy3}) becomes
\begin{equation}\label{Entropy4}
{\rm{d}}S=\frac{1}{T}{\rm{d}}E-\frac{1}{T}(\gamma^{\rm{sp}}-\gamma^{\rm{sl}}-\gamma^{\rm{pl}}){\rm{d}}\sigma^{\rm{sp}}-\frac{1}{T}\sum_{\alpha,\beta}\sum_j\tilde{\mu}_j^{\alpha,\beta}{\rm{d}}n_j^{\alpha,\beta}.
\end{equation}
By equalizing right parts of equations (\ref{Entropy2}) and (\ref{Entropy4}), we obtain after simplifications 
\begin{equation}\label{Entropy5}
{\rm{d}}_iS=\frac{1}{T}\sum_j\tilde{\mu}_j^{\rm{l}}{\rm{d}}_en_j^{\rm{l}}-\frac{1}{T}\sum_{\alpha,\beta}\sum_j\tilde{\mu}_j^{\alpha,\beta}{\rm{d}}n_j^{\alpha,\beta}
\end{equation}
resulting in
\begin{equation}\label{Entropy6}
{\rm{d}}_iS=\frac{1}{T}\sum_j\tilde{\mu}_j^{\rm{l}}{\rm{d}}_en_j^{\rm{l}}-\frac{1}{T}\sum_j\tilde{\mu}_j^{\rm{l}}{\rm{d}}n_j^{\rm{l}}-\frac{1}{T}\sum_{\beta}\sum_j\tilde{\mu}_j^{\beta}{\rm{d}}n_j^{\beta}
\end{equation}
if it is considered that ${\rm{d}}n_j^{\rm{p}}={\rm{d}}n_j^{\rm{s}}=0$ (colloids being considered as closed phases). Variation of the amount of matter contained in the liquid phase is linked to the amount exchanged with the environment of the system but also to amount coming from ``sp'' and ``pl'' superficial phases:
\begin{equation}\label{BilanLiquid}
{\rm{d}}n_j^{\rm{l}}={\rm{d}}_en_j^{\rm{l}}-{\rm{d}}n_j^{\rm{pl}}-{\rm{d}}n_j^{\rm{sl}}.
\end{equation}
Substituting last relation into (\ref{Entropy6}) provides
\begin{equation}\label{Entropy7}
{\rm{d}}_iS=
 \frac{1}{T}\sum_j\tilde{\mu}_j^{\rm{l}}{\rm{d}}n_j^{\rm{pl}}
+\frac{1}{T}\sum_j\tilde{\mu}_j^{\rm{l}}{\rm{d}}n_j^{\rm{sl}}
-\frac{1}{T}\sum_{\beta}\sum_j\tilde{\mu}_j^{\beta}{\rm{d}}n_j^{\beta}
\end{equation}
which simplifies as (\ref{Entropy8}) after development of the last term on $\beta$ and knowing that ${\rm{d}}n_j^{\rm{sp}}=0$, 
\begin{equation}\label{Entropy8}
{\rm{d}}_iS=
\frac{1}{T}\sum_j(\tilde{\mu}_j^{\rm{l}}-\tilde{\mu}_j^{\rm{pl}}){\rm{d}}n_j^{\rm{pl}}
+\frac{1}{T}\sum_j(\tilde{\mu}_j^{\rm{l}}-\tilde{\mu}_j^{\rm{sl}}){\rm{d}}n_j^{\rm{sl}}
\end{equation}

Quantities ${\rm{d}}n_j^{\rm{pl}}$ and ${\rm{d}}n_j^{\rm{sl}}$ are amounts exchanged between liquid phase and interfaces ``pl'' and ``sl''. In other words, they are quantities which are adsorbed or desorbed at/from ``pl'' and ``sl'' interfaces. Adsorption of chemical species $j$ at interface $\beta$ is defined from equation (\ref{DefAdso}) providing the derivative
\begin{equation}
{\rm{d}}n_j^\beta=\Gamma_j^\beta{\rm{d}}\sigma^\beta
\end{equation}
if adsorption $\Gamma_j^\beta$ remains constant during aggregation. Considering equation (\ref{EquivSurfaces}), it comes that
\begin{equation}
{\rm{d}}n_j^{\rm{pl}}=-\Gamma_j^{\rm{pl}}{\rm{d}}\sigma^{\rm{sp}}
\quad\text{and}\quad
{\rm{d}}n_j^{\rm{sl}}=-\Gamma_j^{\rm{sl}}{\rm{d}}\sigma^{\rm{sp}}
\end{equation}
providing equation (\ref{Entropy9}) after substitution into (\ref{Entropy8}) and rearrangement.
\begin{equation}\label{Entropy9}
{\rm{d}}_iS=
\frac{1}{T}\sum_j(\tilde{\mu}_j^{\rm{pl}}-\tilde{\mu}_j^{\rm{l}})\Gamma_j^{\rm{pl}}{\rm{d}}\sigma^{\rm{sp}}
+\frac{1}{T}\sum_j(\tilde{\mu}_j^{\rm{sl}}-\tilde{\mu}_j^{\rm{l}})\Gamma_j^{\rm{sl}}{\rm{d}}\sigma^{\rm{sp}}
\end{equation}

Equation (\ref{Entropy9}) is the final relation allowing to describe the out-of-equilibrium aggregation process. When injected into (\ref{Entropy2}), it provides the full expression for entropy derivative during a time interval ${\rm{d}}t$ of an aggregation process and will make sense in the following discussion.
\end{widetext}

\section{Discussion}\label{ThirdSection}
\subsection{Source of irreversibility}

We have seen that ${\rm{d}}_iS\geqslant0$ is a condition for natural processes. Natural means spontaneous and irreversible \citep{planck1913,guggenheim1949}. The second law gives thus the direction of spontaneous evolution of state variables describing the out-of-equilibrium system \citep{prigogine1968}. In the present case, it will answer the existential question: Do colloids ``s'' and ``p'' spontaneously aggregate or not?

Aggregation of ``s'' and ``p'' colloids means that they stick together but do not merge, the latter being coalescence. The consequence of aggregation is the increase of contact area between the two colloids; that is, ${\rm{d}}\sigma^{\rm{sp}}>0$. If ${\rm{d}}\sigma^{\rm{sp}}>0$, aggregation is processing and if ${\rm{d}}\sigma^{\rm{sp}}=0$, the system has reached a stable state, but not necessarily equilibrium (see below). Conversely, a system for which ${\rm{d}}\sigma^{\rm{sp}}<0$ is said to be peptizing \citep{verwey1948}.

During these processes, adsorptions $\Gamma_j^\beta$ remain constant because they are intensive variables. For species in excess at the interface ($\Gamma_j^\beta>0$), aggregation requires that $\tilde{\mu}_j^\beta>\tilde{\mu}_j^{\rm{l}}$ to fulfil the second law; that is, ${\rm{d}}_iS\geqslant0$. For $\Gamma_j^\beta<0$, we have the opposite: $\tilde{\mu}_j^\beta<\tilde{\mu}_j^{\rm{l}}$. If these conditions are met, then the aggregation of colloids may provoke desorption and in consequence, entropy production.

Desorption is not the only entropic source. Desorption solely describes the release of chemical species from superficial phases. After their release, high concentrations could appear near the space region where desorption took place. These conditions characterized by concentration gradients are supposed to create a diffusional flow toward the region of lower concentration; that is, the core of the bulk phase.

Fick's second law describes the diffusion of chemical species against their respective concentration gradients. This law is phenomenological inasmuch as it only describes what is happening; that is, species migrate along their gradients. Fick's second law does not describe their cause of diffusion; that is, why chemical species migrate. The cause of migration is the minimization of chemical potentials \citep{prigogine1968,lebon2008a} thus molecules will migrate against their chemical potential gradient. A chemical species $j$ will migrate from place I to place II if its chemical potential (related to activity and thus concentration) in place II is weaker than in place I, a condition we write as $\mu_j^\mathrm{II}<\mu_j^\mathrm{I}$. Lowering chemical potentials is a source of energy dissipation and thus of entropic production \citep{lebon2008a} as equation (\ref{EntropyDiff}) shows.
\begin{equation}\label{EntropyDiff}
{\rm{d}}_iS_\mathrm{diff}=-\frac{1}{T}\int_v \sum_j J_j \nabla \tilde{\mu}_j {\rm{d}}v
\end{equation}

As a consequence, for species showing positive adsorption at interfaces ($\Gamma_j^\beta>0$), spontaneous aggregation will induce desorption from the portions of interfaces becoming contact area between colloids, and the diffusion of these desorbed species toward the core of the suspending liquid phase.

\subsection{De Donder's affinities for adsorption and desorption}

Affinities are quantities closely linked to entropy production and consequently widely used when dealing with out-of-equilibrium physico-chemical process \citep{prigogine1950}. They have been introduced by Th. De Donder \citep{prigogine1968} on the basis of chemical potentials. Adsorption affinities of $j$ at ``pl'' and ``sl'' interfaces are definied such that \citep{defay1951,defay1977}
\begin{equation}
\mathcal{A}_j^{\rm{pl}}=\mu_j^{\rm{l}}-\mu_j^{\rm{pl}}
\quad\text{and}\quad
\mathcal{A}_j^{\rm{sl}}=\mu_j^{\rm{l}}-\mu_j^{\rm{sl}}.
\end{equation}
At constant $T$ and $p$, affinities are free enthalpy released by the adsorption of one mole of $j$ at ``pl'' or ``sl'' interfaces. In the same way, desorption affinities may be oppositely defined by
\begin{equation}\label{DesorpAff}
-\mathcal{A}_j^{\rm{pl}}=\mu_j^{\rm{pl}}-\mu_j^{\rm{l}}
\quad\text{and}\quad
-\mathcal{A}_j^{\rm{sl}}=\mu_j^{\rm{sl}}-\mu_j^{\rm{l}}.
\end{equation}

If the electrochemical desorption affinities are defined on the basis of electrochemical potentials instead of chemical potentials, it follows
\begin{equation}
-\tilde{\mathcal{A}}_j^{\rm{pl}}=-\mathcal{A}_j^{\rm{pl}}+z_j\mathcal{F}\psi^{\rm{pl}}
\quad\text{and}\quad
-\tilde{\mathcal{A}}_j^{\rm{sl}}=-\mathcal{A}_j^{\rm{sl}}+z_j\mathcal{F}\psi^{\rm{sl}}
\end{equation}
 from (\ref{DesorpAff}), (\ref{ChemicalPotential}), and the fact that $\psi^{\rm{l}}=0$.
 
Introducing electrochemical desorption affinities into equation (\ref{Entropy9}) provides
\begin{equation}\label{Entropy10}
{\rm{d}}_iS=
-\frac{1}{T}\sum_j\tilde{\mathcal{A}}_j^{\rm{pl}}\Gamma_j^{\rm{pl}}{\rm{d}}\sigma^{\rm{sp}}
-\frac{1}{T}\sum_j\tilde{\mathcal{A}}_j^{\rm{sl}}\Gamma_j^{\rm{sl}}{\rm{d}}\sigma^{\rm{sp}}.
\end{equation}
Affinities are useful to find out if the system is an equilibrated one. Positive affinity means that the system is out-of-equilibrium and may produce entropy. Moreover, from de Gibbs' conditions of equilibrium (\ref{GibbsEquili}), the system will be considered in an equilibrium state if and only if all the affinities are zero:
\begin{equation}
\tilde{\mathcal{A}}_j^{\rm{pl}}=\dots =\tilde{\mathcal{A}}_j^{\rm{sl}}=\dots =0.
\end{equation}

A system which has been moved away from equilibrium; that is, for which affinities are positive, will tend to equilibrium by nullifying all the affinities. When affinities reach zero, the evolution process will end, and entropy production will stop. It is worth noting that equilibrium conditions are not the same as stability conditions. Indeed, during the evolution toward equilibrium, stability conditions can be met which would stop the process before the equilibrium conditions have been reached. Conversely, equilibrium is necessarily a stable state.

Thus, a stable (no entropy production) colloidal system may stay out of equilibrium (positive affinities). Breaking one or more stability conditions makes possible evolution toward equilibrium with an entropy production. Aggregation requires desorption to produce entropy but happens only if the stability conditions of the system have been broken.

\subsection{Colloidal (in)stability}

A colloidal system is stabilized by summing different well known potentials \citep{verwey1948}: the Hamaker and electrostatic potentials. Hamaker's potential results from the summation of London-van der Waals potential between each portion of interacting bodies \citep{hamaker1937}. It is often attractive but for two colloids of the same nature, say ``p'' and ``p'' or ``s'' and ``s'', it is violently attractive at short distance. Electrostatic potential can be attractive or repulsive depending on superficial charges borne by the colloids. Its intensity depends on the nature of colloidal surfaces and on physico-chemical conditions in the liquid phase, \textit{i.e.} pH, ionic forces, \textit{etc}. For two colloids of the same nature and uniformly charged, it is automatically repulsive \citep{verwey1948}.

Summing those two potentials provides the DLVO model, a resulting potential steering the stability of our colloidal system \citep{verwey1948}. Indeed, these deduced potential could show energy barrier preventing an approach between colloids, but if underlying potentials depend on physico-chemical conditions, so does the energy barrier. Therefore, lowering the DLVO potential energy barrier could destabilize the system because of a decrease in the Debye length \citep{verwey1948}. The Debye length decrease causes an increasing of the probability that two colloids closely approach and then fall into the trap of strongly attractive Hamaker's potential.

When two colloids ``p'' and ``p'' (or ``s'' and ``s'') approach so that Hamaker's potential is not thwarted, it will result on compression of superficial layers causing an increase of $\tilde{\mu}_j^{\rm{pl}}$ (or $\tilde{\mu}_j^{\rm{sl}}$). Such an increase of superficial chemical potentials also produces an increase of electrochemical desorption affinities $-\tilde{\mathcal{A}}_j^{\rm{pl}}$ and $-\tilde{\mathcal{A}}_j^{\rm{sl}}$ which is the indication that the system is moved away from equilibrium.

To return toward the equilibrium (lowering affinities), desorption occurs spontaneously because ${\rm{d}}_iS>0$. Desorption occurs until stability has been reached. This phenomena where two colloids of the same nature aggregate by banishing solvent and cosolvent molecules from interspaces between each other to produce entropy may most likely be compared to the hydrophobic effect. This hydrophobic effect tends to increase contact area between two bodies such as macromolecules dispersed into hostile water environment.

In this way, a distinction between adsorption and aggregation is made. The first occurring for low molecular weight molecules (water, salts, \textit{etc}.) and the second for macromolecular bodies such as colloids (proteins, polymers, \textit{etc.}). As an example, a protein could aggregate on a hydrophobic surface because of the possibility of low molecular weight molecules desorption from their surfaces.

\section*{Conclusion}

With the aim to reach a better understanding of hydrophobic effect, equations describing the aggregation of two hydrophobic colloids have been shown. This description is based on purely thermodynamic considerations and specifically on classical irreversible thermodynamics. In fact, according to the second law, systems which transform are characterized by entropy production until they recover spontaneously and irreversibly a stable state not necessarily at equilibrium.

In equations (\ref{Entropy9}) and (\ref{Entropy10}), entropy production (${\rm{d}}_iS$) is linked to the contact area variation (${\rm{d}}\sigma^{\rm{sp}}$) between colloids ``s'' and ``p''. Proportionality factors are De Donder's affinities of desorption from the surfaces of the two aggregating colloids and the adsorbed quantities. These equations brings to a new explanation for hydrophobic effect: suppose the initial stability conditions of the system were broken, this would allow colloids to approach beyond the potential energy barrier (see DLVO theory). Hamaker's potential may then violently attract the two colloids. In turn this attraction provokes an increase of desorption affinities (because chemical potentials of superficial species increase) allowing adsorbed species to desorb and produce entropy. Due to entropy production, aggregation is then irreversible and spontaneous. A second entropic source from diffusion has also been identified.

This development shows that the solvent (water) and cosolvent (ions, \textit{etc}.) molecules decisively contribute to hydrophobic interaction. Furthermore, equations use chemical potentials of these molecules but do not for colloids. Indeed, defining chemical potential for a colloid (macromelcules, proteins, \textit{etc}.) does not appear to be a judicious idea.

In addition to the use of non-equilibrium thermodynamics, equations (\ref{Entropy9}) and (\ref{Entropy10}) are evidences that superficial events could be of major interest for biological interactions. They offer real perspectives in the understanding of colloidal thermodynamics but also of protein interactions, their stability, and folding. In fact, we could well imagine that colloid ``s'' is the variable part of an antibody, and that the colloid ``p'' is an antigen, the first recognizing the second.

\begin{acknowledgments}
The author has received funds from R\'{e}gion wallonne (Belgium) for 2.5 years and would like also thanks F\'{e}lix de Thier for valuable discussion.
\end{acknowledgments}


\begin{thebibliography}{23}%
\makeatletter
\providecommand \@ifxundefined [1]{%
 \@ifx{#1\undefined}
}%
\providecommand \@ifnum [1]{%
 \ifnum #1\expandafter \@firstoftwo
 \else \expandafter \@secondoftwo
 \fi
}%
\providecommand \@ifx [1]{%
 \ifx #1\expandafter \@firstoftwo
 \else \expandafter \@secondoftwo
 \fi
}%
\providecommand \natexlab [1]{#1}%
\providecommand \enquote  [1]{``#1''}%
\providecommand \bibnamefont  [1]{#1}%
\providecommand \bibfnamefont [1]{#1}%
\providecommand \citenamefont [1]{#1}%
\providecommand \href@noop [0]{\@secondoftwo}%
\providecommand \href [0]{\begingroup \@sanitize@url \@href}%
\providecommand \@href[1]{\@@startlink{#1}\@@href}%
\providecommand \@@href[1]{\endgroup#1\@@endlink}%
\providecommand \@sanitize@url [0]{\catcode `\\12\catcode `\$12\catcode
  `\&12\catcode `\#12\catcode `\^12\catcode `\_12\catcode `\%12\relax}%
\providecommand \@@startlink[1]{}%
\providecommand \@@endlink[0]{}%
\providecommand \url  [0]{\begingroup\@sanitize@url \@url }%
\providecommand \@url [1]{\endgroup\@href {#1}{\urlprefix }}%
\providecommand \urlprefix  [0]{URL }%
\providecommand \Eprint [0]{\href }%
\providecommand \doibase [0]{http://dx.doi.org/}%
\providecommand \selectlanguage [0]{\@gobble}%
\providecommand \bibinfo  [0]{\@secondoftwo}%
\providecommand \bibfield  [0]{\@secondoftwo}%
\providecommand \translation [1]{[#1]}%
\providecommand \BibitemOpen [0]{}%
\providecommand \bibitemStop [0]{}%
\providecommand \bibitemNoStop [0]{.\EOS\space}%
\providecommand \EOS [0]{\spacefactor3000\relax}%
\providecommand \BibitemShut  [1]{\csname bibitem#1\endcsname}%
\let\auto@bib@innerbib\@empty
%</preamble>
\bibitem [{\citenamefont {Chandler}(2005)}]{chandler2005}%
  \BibitemOpen
  \bibfield  {author} {\bibinfo {author} {\bibfnamefont {D.}~\bibnamefont
  {Chandler}},\ }\href@noop {} {\bibfield  {journal} {\bibinfo  {journal}
  {Nature}\ }\textbf {\bibinfo {volume} {437}},\ \bibinfo {pages} {640}
  (\bibinfo {year} {2005})}\BibitemShut {NoStop}%
\bibitem [{\citenamefont {Tanford}(1978)}]{tanford1978}%
  \BibitemOpen
  \bibfield  {author} {\bibinfo {author} {\bibfnamefont {C.}~\bibnamefont
  {Tanford}},\ }\href@noop {} {\bibfield  {journal} {\bibinfo  {journal}
  {Science}\ }\textbf {\bibinfo {volume} {200}},\ \bibinfo {pages} {1012}
  (\bibinfo {year} {1978})}\BibitemShut {NoStop}%
\bibitem [{\citenamefont {Levy}\ and\ \citenamefont
  {Onuchic}(2006)}]{levy2006}%
  \BibitemOpen
  \bibfield  {author} {\bibinfo {author} {\bibfnamefont {Y.}~\bibnamefont
  {Levy}}\ and\ \bibinfo {author} {\bibfnamefont {J.}~\bibnamefont {Onuchic}},\
  }\href@noop {} {\bibfield  {journal} {\bibinfo  {journal} {Annual Review of
  Biophysics and Biomolecular Structure}\ }\textbf {\bibinfo {volume} {35}},\
  \bibinfo {pages} {389} (\bibinfo {year} {2006})}\BibitemShut {NoStop}%
\bibitem [{\citenamefont {Nelson}\ and\ \citenamefont {Cox}(2005)}]{lehninger}%
  \BibitemOpen
  \bibfield  {author} {\bibinfo {author} {\bibfnamefont {D.}~\bibnamefont
  {Nelson}}\ and\ \bibinfo {author} {\bibfnamefont {M.}~\bibnamefont {Cox}},\
  }\href@noop {} {\emph {\bibinfo {title} {Lehninger Principles of
  Biochemistry. Fourth edition.}}}\ (\bibinfo  {publisher} {W.H. Freeman and
  Company},\ \bibinfo {address} {New York},\ \bibinfo {year}
  {2005})\BibitemShut {NoStop}%
\bibitem [{\citenamefont {Dill}\ and\ \citenamefont
  {MacCallum}(2012)}]{dill2012}%
  \BibitemOpen
  \bibfield  {author} {\bibinfo {author} {\bibfnamefont {K.}~\bibnamefont
  {Dill}}\ and\ \bibinfo {author} {\bibfnamefont {J.}~\bibnamefont
  {MacCallum}},\ }\href@noop {} {\bibfield  {journal} {\bibinfo  {journal}
  {Science}\ }\textbf {\bibinfo {volume} {338}},\ \bibinfo {pages} {1042}
  (\bibinfo {year} {2012})}\BibitemShut {NoStop}%
\bibitem [{\citenamefont {Brady}\ and\ \citenamefont
  {Sharp}(1997)}]{brady1997}%
  \BibitemOpen
  \bibfield  {author} {\bibinfo {author} {\bibfnamefont {G.}~\bibnamefont
  {Brady}}\ and\ \bibinfo {author} {\bibfnamefont {K.}~\bibnamefont {Sharp}},\
  }\href@noop {} {\bibfield  {journal} {\bibinfo  {journal} {Current Opinion in
  Structural Biology}\ }\textbf {\bibinfo {volume} {7}},\ \bibinfo {pages}
  {215} (\bibinfo {year} {1997})}\BibitemShut {NoStop}%
\bibitem [{\citenamefont {Lum}, \citenamefont {Chandler},\ and\ \citenamefont
  {Weeks}(1999)}]{lum1999}%
  \BibitemOpen
  \bibfield  {author} {\bibinfo {author} {\bibfnamefont {K.}~\bibnamefont
  {Lum}}, \bibinfo {author} {\bibfnamefont {D.}~\bibnamefont {Chandler}}, \
  and\ \bibinfo {author} {\bibfnamefont {J.}~\bibnamefont {Weeks}},\
  }\href@noop {} {\bibfield  {journal} {\bibinfo  {journal} {Journal of
  Physical Chemistry B}\ }\textbf {\bibinfo {volume} {103}},\ \bibinfo {pages}
  {4570} (\bibinfo {year} {1999})}\BibitemShut {NoStop}%
\bibitem [{\citenamefont {Verwey}\ and\ \citenamefont
  {Overbeek}(1948)}]{verwey1948}%
  \BibitemOpen
  \bibfield  {author} {\bibinfo {author} {\bibfnamefont {E.}~\bibnamefont
  {Verwey}}\ and\ \bibinfo {author} {\bibfnamefont {J.}~\bibnamefont
  {Overbeek}},\ }\href@noop {} {\emph {\bibinfo {title} {Theory of the
  stability of lyophobic colloids}}}\ (\bibinfo  {publisher} {Elsevier
  Publishing Company, Inc.},\ \bibinfo {address} {New York},\ \bibinfo {year}
  {1948})\BibitemShut {NoStop}%
\bibitem [{\citenamefont {Schr\"{o}dinger}(1967)}]{schrodinger1967}%
  \BibitemOpen
  \bibfield  {author} {\bibinfo {author} {\bibfnamefont {E.}~\bibnamefont
  {Schr\"{o}dinger}},\ }\href@noop {} {\emph {\bibinfo {title} {What is
  Life?}}}\ (\bibinfo  {publisher} {Cambridge University Press},\ \bibinfo
  {address} {London},\ \bibinfo {year} {1967})\BibitemShut {NoStop}%
\bibitem [{\citenamefont {Planck}(1913)}]{planck1913}%
  \BibitemOpen
  \bibfield  {author} {\bibinfo {author} {\bibfnamefont {M.}~\bibnamefont
  {Planck}},\ }\href@noop {} {\emph {\bibinfo {title} {Le\c{c}ons de
  Thermodynamique}}},\ \bibinfo {edition} {ouvrage traduit sur le troisi\`{e}me
  \'{e}dition allemande (augment\'{e}e)}\ ed.\ (\bibinfo  {publisher}
  {Librairie Scientifique A. Hermann et Fils},\ \bibinfo {address} {Paris},\
  \bibinfo {year} {1913})\BibitemShut {NoStop}%
\bibitem [{\citenamefont {Guggenheim}(1949)}]{guggenheim1949}%
  \BibitemOpen
  \bibfield  {author} {\bibinfo {author} {\bibfnamefont {E.}~\bibnamefont
  {Guggenheim}},\ }\href@noop {} {\emph {\bibinfo {title} {Thermodynamics}}},\
  Monographs on theoretical and applied physics\ (\bibinfo  {publisher}
  {North-Holland Publishing Company},\ \bibinfo {address} {Amsterdam},\
  \bibinfo {year} {1949})\BibitemShut {NoStop}%
\bibitem [{\citenamefont {Defay}, \citenamefont {Prigogine},\ and\
  \citenamefont {Sanfeld}(1977)}]{defay1977}%
  \BibitemOpen
  \bibfield  {author} {\bibinfo {author} {\bibfnamefont {R.}~\bibnamefont
  {Defay}}, \bibinfo {author} {\bibfnamefont {I.}~\bibnamefont {Prigogine}}, \
  and\ \bibinfo {author} {\bibfnamefont {A.}~\bibnamefont {Sanfeld}},\
  }\href@noop {} {\bibfield  {journal} {\bibinfo  {journal} {Journal of Colloid
  and Interface Science}\ }\textbf {\bibinfo {volume} {58}},\ \bibinfo {pages}
  {498} (\bibinfo {year} {1977})}\BibitemShut {NoStop}%
\bibitem [{\citenamefont {Prigogine}(1968)}]{prigogine1968}%
  \BibitemOpen
  \bibfield  {author} {\bibinfo {author} {\bibfnamefont {I.}~\bibnamefont
  {Prigogine}},\ }\href@noop {} {\emph {\bibinfo {title} {Introduction \`{a} la
  thermodynamique des processus irr\'{e}versibles}}}\ (\bibinfo  {publisher}
  {Dunod},\ \bibinfo {address} {Paris},\ \bibinfo {year} {1968})\BibitemShut
  {NoStop}%
\bibitem [{\citenamefont {Reiss}(1965)}]{reiss1965}%
  \BibitemOpen
  \bibfield  {author} {\bibinfo {author} {\bibfnamefont {H.}~\bibnamefont
  {Reiss}},\ }\href@noop {} {\emph {\bibinfo {title} {Methods of
  thermodynamics}}}\ (\bibinfo  {publisher} {Blaisdell Publishing Company},\
  \bibinfo {address} {New York},\ \bibinfo {year} {1965})\BibitemShut {NoStop}%
\bibitem [{\citenamefont {Nagarajan}\ and\ \citenamefont
  {Ruckenstein}(1991)}]{nagarajan1991}%
  \BibitemOpen
  \bibfield  {author} {\bibinfo {author} {\bibfnamefont {R.}~\bibnamefont
  {Nagarajan}}\ and\ \bibinfo {author} {\bibfnamefont {E.}~\bibnamefont
  {Ruckenstein}},\ }\href@noop {} {\bibfield  {journal} {\bibinfo  {journal}
  {Langmuir}\ }\textbf {\bibinfo {volume} {7}},\ \bibinfo {pages} {2934}
  (\bibinfo {year} {1991})}\BibitemShut {NoStop}%
\bibitem [{\citenamefont {Prigogine}\ and\ \citenamefont
  {Defay}(1950)}]{prigogine1950}%
  \BibitemOpen
  \bibfield  {author} {\bibinfo {author} {\bibfnamefont {I.}~\bibnamefont
  {Prigogine}}\ and\ \bibinfo {author} {\bibfnamefont {R.}~\bibnamefont
  {Defay}},\ }\href@noop {} {\emph {\bibinfo {title} {Thermodynamique
  Chimique}}},\ \bibinfo {series} {Trait\'{e} de Thermodynamique
  conform\'{e}ment aux m\'{e}thodes de Gibbs et De Donder}, Vol.\ \bibinfo
  {volume} {I et II}\ (\bibinfo  {publisher} {Editions Desoer},\ \bibinfo
  {address} {Li\`{e}ge},\ \bibinfo {year} {1950})\BibitemShut {NoStop}%
\bibitem [{\citenamefont {Defay}\ and\ \citenamefont
  {Prigogine}(1951)}]{defay1951}%
  \BibitemOpen
  \bibfield  {author} {\bibinfo {author} {\bibfnamefont {R.}~\bibnamefont
  {Defay}}\ and\ \bibinfo {author} {\bibfnamefont {I.}~\bibnamefont
  {Prigogine}},\ }\href@noop {} {\emph {\bibinfo {title} {Tension superficielle
  et adsorption}}},\ \bibinfo {series} {Trait\'{e} de Thermodynamique
  conform\'{e}ment aux m\'{e}thodes de Gibbs et De Donder}, Vol.\ \bibinfo
  {volume} {III}\ (\bibinfo  {publisher} {Edition Desoer},\ \bibinfo {address}
  {Li\`{e}ge},\ \bibinfo {year} {1951})\BibitemShut {NoStop}%
\bibitem [{\citenamefont {Hunter}(1987)}]{hunter1987}%
  \BibitemOpen
  \bibfield  {author} {\bibinfo {author} {\bibfnamefont {R.}~\bibnamefont
  {Hunter}},\ }\href@noop {} {\emph {\bibinfo {title} {Foundations of Colloid
  Science}}},\ \bibinfo {series} {Oxford Science Publications}, Vol.~\bibinfo
  {volume} {I}\ (\bibinfo  {publisher} {Oxford University Press},\ \bibinfo
  {address} {New York},\ \bibinfo {year} {1987})\BibitemShut {NoStop}%
\bibitem [{Note1()}]{Note1}%
  \BibitemOpen
  \bibinfo {note} {It is customary to define chemical potential as the partial
  derivative of free enthalpy at $p$ and $T$ constant. Here, for the sake of
  simplicity, it is better to use the derivative of $E$ keeping $S$, $v$ and
  $\sigma $ constant.}\BibitemShut {Stop}%
\bibitem [{\citenamefont {Lyklema}(1991)}]{lyklema1}%
  \BibitemOpen
  \bibfield  {author} {\bibinfo {author} {\bibfnamefont {H.}~\bibnamefont
  {Lyklema}},\ }\href@noop {} {\emph {\bibinfo {title} {Fundamentals}}},\
  \bibinfo {series} {Fundamentals of Interfaces and Colloid Science},
  Vol.~\bibinfo {volume} {I}\ (\bibinfo  {publisher} {Academic Press},\
  \bibinfo {address} {London},\ \bibinfo {year} {1991})\BibitemShut {NoStop}%
\bibitem [{\citenamefont {Atkins}\ and\ \citenamefont
  {De~Paula}(2006)}]{atkins}%
  \BibitemOpen
  \bibfield  {author} {\bibinfo {author} {\bibfnamefont {P.}~\bibnamefont
  {Atkins}}\ and\ \bibinfo {author} {\bibfnamefont {J.}~\bibnamefont
  {De~Paula}},\ }\href@noop {} {\emph {\bibinfo {title} {Physical Chemistry}}}\
  (\bibinfo  {publisher} {Oxford University Press},\ \bibinfo {address}
  {Oxford},\ \bibinfo {year} {2006})\BibitemShut {NoStop}%
\bibitem [{\citenamefont {Lebon}, \citenamefont {Jou},\ and\ \citenamefont
  {Casas-V\'{a}zquez}(2008)}]{lebon2008a}%
  \BibitemOpen
  \bibfield  {author} {\bibinfo {author} {\bibfnamefont {G.}~\bibnamefont
  {Lebon}}, \bibinfo {author} {\bibfnamefont {D.}~\bibnamefont {Jou}}, \ and\
  \bibinfo {author} {\bibfnamefont {J.}~\bibnamefont {Casas-V\'{a}zquez}},\
  }in\ \href@noop {} {\emph {\bibinfo {booktitle} {Understanding
  Non-equilibrium Thermodynamics}}}\ (\bibinfo  {publisher} {Springer Berlin
  Heidelberg},\ \bibinfo {year} {2008})\ pp.\ \bibinfo {pages}
  {1--36}\BibitemShut {NoStop}%
\bibitem [{\citenamefont {Hamaker}(1937)}]{hamaker1937}%
  \BibitemOpen
  \bibfield  {author} {\bibinfo {author} {\bibfnamefont {H.}~\bibnamefont
  {Hamaker}},\ }\href@noop {} {\bibfield  {journal} {\bibinfo  {journal}
  {Physica}\ }\textbf {\bibinfo {volume} {4}},\ \bibinfo {pages} {1058}
  (\bibinfo {year} {1937})}\BibitemShut {NoStop}%
\end{thebibliography}%
\end{document}